\documentclass[review]{elsarticle}

\usepackage{lineno,hyperref}
\modulolinenumbers[5]
\usepackage{graphicx}
\usepackage[numbers]{natbib}
\usepackage[center]{caption}
\usepackage{amsmath}
\usepackage[ruled,vlined]{algorithm2e}

\begin{document}

\begin{frontmatter}

\title{An Efficient Lightweight Blockchain for Decentralized IoT}

\author{Faezeh Dehghan}
\ead{Faezehdehghan@ut.ac.ir}

\author{Mostafa Salehi\corref{mycorrespondingauthor}}
\cortext[mycorrespondingauthor]{Corresponding author}
\ead{Mostafa_salehi@ut.ac.ir}

\address{University of Tehran,Tehran, Iran}

\begin{abstract}
The Internet of Things (IoT) is applied in various fields, and the number of physical devices connected to the IoT is increasingly growing. There are significant challenges to the IoT's growth and development, mainly due to the centralized nature and large-scale IoT networks. The emphasis on the decentralization of IoT's architecture can overcome challenges to IoT's capabilities. A promising decentralized platform for IoT is blockchain. Owing to IoT devices' limited resources, traditional consensus algorithms such as PoW and PoS in the blockchain are computationally expensive. Therefore, the PoA consensus algorithm is proposed in the blockchain consensus network for IoT. The PoA selects the validator as Turn-based selection (TBS) that needs optimization and faces system reliability, energy consumption, latency, and low scalability. We propose an efficient, lightweight blockchain for decentralizing IoT architecture by using virtualization and clustering to increase productivity and scalability to address these issues. We also introduce a novel PoA based on the Weight-Based-Selection (WBS) method for validators to validate transactions and add them to the blockchain. By simulation, we evaluated the performance of our proposed WBS method as opposed to TBS. The results show reduced energy consumption, and response time, and increased throughput.
\end{abstract}


\end{frontmatter}

\section{Introduction}
Internet of Things (IoT) is composed of objects that can be linked via communication networks, with the ability to exchange information to improve intelligence, service quality, positioning, monitoring, control, and security \cite{zhang2014iot}. In recent years, we have witnessed the potential of IoT to provide services in a range of industries, including cities, homes, transportation, industrial factories, agriculture environments, healthcare centers, and smart energy \cite{lee2015internet}.
\par
\begin{figure*}
 \center
  \includegraphics[width=1.02\textwidth]{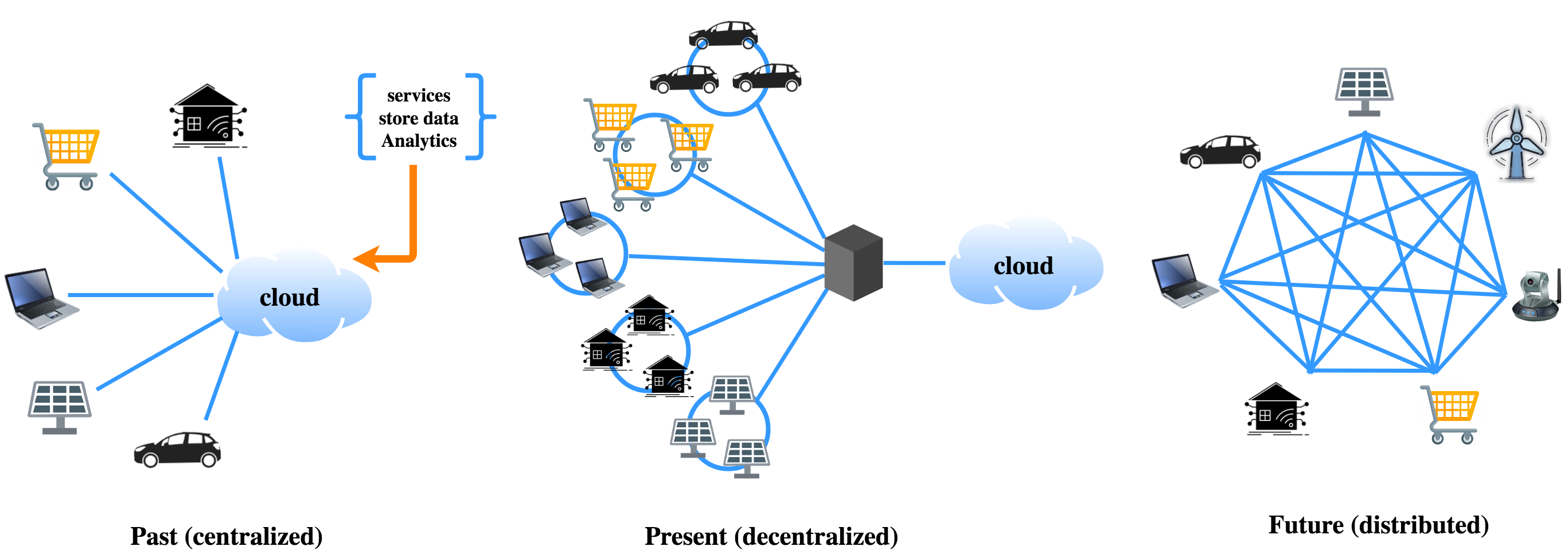}
  \caption{The past, present, and future of IoT architecture.}
  \label{ppf}
\end{figure*}

The number of physical devices connected to the IoT is increasingly growing \cite{sharma2019history}. IDC predicts that there seem to be about 41.6 billion connected objects in 2025 \cite{wasicek2020future}. As shown in \figurename{\ref{ppf}}, The IoT architecture used to be cloud-based with centralized access. This architecture is then replaced by placing a middle layer (fog nodes) next to the centralized cloud to conduct simple processing and minimize latency by being close to the end node geographically \cite{chiang2016fog}. Most communications and IoT services, however, still depend on these communication models \cite{hassija2019survey}. High maintenance costs, weak interoperability, single point of failure (SPOF) against security threats, low scalability, many-to-one traffic, and a lack of prompt support are all weaknesses of this centralized model \cite{roman2013features}. Given the increase in both numbers and requirements of IoT-connected devices, in the future, a decentralized, secure, robust approach is expected in which cloud functionality is distributed among nodes; and objects via the Internet can communicate directly with others \cite{lee2017future}.
\par
 Decentralized IoT architecture, while having advantages, faces particular security challenges such as managing connections between heterogeneous devices, identifying, authenticating, managing devices, access-level permissions, secure storage of information in nodes, and preventing data release to various cyber-attacks \cite{sicari2015security}.
 \par A possible decentralized platform for IoT is blockchain. Blockchain is the concept of a distributed immutable ledger maintained via a peer-to-peer network \cite{khan2018iot}. Recent research presents a wide variety of blockchain potential in the IoT, applications such as secure communication between devices through smart contracts, secure data storage, device administration, and authentication, access control, privacy improvements, and third-party removal \cite{hassija2019survey,latif2022ai,fotohi2021securing}.

\par Early research on the blockchain in the IoT offered classic consensus algorithms such as Proof of Work (PoW) \cite{gervais2016security} and Proof of Stack (PoS) \cite{saleh2021blockchain}, which are computationally demanding owing to the complicated mathematical issue (generating input of a hash function) and are not suited for IoT devices with low processing power. Also, the long transaction authentication time did not support the issue of IoT scalability. After introducing the proof of authority (PoA) \cite{de2018pbft} consensus algorithm in 2017, various studies were performed to develop a lightweight blockchain leveraging the PoA consensus algorithm following the IoT ecosystem to resolve this challenge. This algorithm's advantages over the classic consensus algorithm are low power consumption, high throughput, and low latency \cite{puthal2018proof}.
\par 
The number of researchers adopting blockchain in the IoT has increased in recent years, and significant advances have been reached \cite{uddin2021survey}, yet there are still challenges hidden from researchers. Resource management in distributed environments such as blockchain is more important due to more workload, dynamic and interdependent requests. Furthermore, one of the issues of decentralizing IoT architecture is node energy management \cite{arshad2017green}. The more nodes interact in the network, the more crucial it becomes to find a solution to this challenge. Inefficient utilization of available resources is one of the reasons for energy loss.
\par 
The PoA consensus algorithm is utilized as a Turn-based Selection (TBS); a validator is selected in each round to validate transactions and add a block to the blockchain. If the selected validator fails and is not available at the indicated time, the requested service cannot be guaranteed. Also, when the number of requests is high, or the network is experiencing high traffic, the given validator may not be the best choice to accomplish tasks at the specified time. Consequently, the request-response time may increase, and the network may become overloaded due to processing delays.
\par Another challenge with blockchain integration in the IoT is scalability \cite{hassan2019privacy}. The amount and speed of data production and processing in the IoT are considerable; increased traffic and information density in the system causes delays and reduces system performance. Algorithms designed for processing across nodes must be scalable according to request and demand. System reliability, service downtime, and accessibility are also issues with decentralized systems.
\par
To improve the usage of blockchain in the IoT, we suggest solutions inspired by cloud computing \cite{javadpour2018power}.\\ Clustering is one technique to reduce energy consumption and enhance scalability in IoT wireless sensor networks. In this approach, nodes are grouped into clusters so that high-energy nodes are used in the head cluster role to process and send data. Sensor nodes send their data alternately to their head clusters. Head cluster nodes collect and process data and send a transaction over the consensus network if a transaction needs to be created. Validators receive these transactions and check their validity. Once the transaction is approved, permission is granted to execute it or record it in the blockchain. This reduces the sensor nodes' communication overhead, which sends data directly to the validator, increasing network life, reducing power consumption, and increasing scalability.\\
Another potential method is the virtualization of the validator's physical machine. Virtual machines are separate, isolated software on which the blockchain application is located. Virtual machines run on the validators' physical machines in the blockchain consensus network. These applications are responsible for verifying transactions and executing smart contracts.
Virtualization makes better use of resources, which reduces costs and energy usage. Virtualization increases scalability and reduces response time by dividing tasks and enhancing safety and more pleasant management of services.
\\
We introduced selecting a validator to validate transactions and add them to the blockchain method of selecting a validator based on weight (WBS). A weight is assigned to each available validator in this method, suggesting that the validator's processing capacity and the validator with a superior degree of attractiveness are selected to fulfill the tasks. 
\par To evaluate the proposed method, we simulated system architecture using the NS2 emulator \cite{issariyakul2009introduction}, a widely used simulator in IoT. We used the MannaSim \cite{pereira2015mannasim} framework to implement the IoT WSN and the Green-Cloud \cite{kliazovich2012greencloud} framework to implement the blockchain network. We have considered the criteria response time, throughput, and energy consumption to evaluate our proposed method. Also, because of various IoT applications and different tasks of devices, we have considered different transaction dissemination intervals from IoT sensors and the different number of validators in the blockchain as variable parameters. The simulation results show that we have improved the blockchain network in our proposed method in terms of reducing energy consumption and response time and increasing throughput.
\par The contribution of this study can be summarized in the following cases:\\
\begin{itemize}
  \item Proposed framework based on clustering and virtualization in using blockchain for decentralized IoT architecture.
  \item Introduce the weight-based-Selection method (WBS) which utilize the attractiveness variable for weighting validators in the blockchain PoA consensus network.
  \item Improving the criteria for reducing energy consumption, increasing scalability and throughput, and reducing response time compared in WBS to the TBS method by selecting a proper validator with weighting based on attractiveness variable.
\end{itemize}
\par
This article's remainder is structured as follows: Section II presents a brief background on the blockchain. In Section III, some related studies are reviewed. In section IV, we detailed our system architecture. In section V we proposed our simulation procedure and simulation configurations. In Section VI, the results of the implementation and evaluation of the proposed method are reported. Finally, in section VII, we draw our conclusions and future studies.
\section{Preliminaries}
Blockchain has numerous benefits, such as decentralization,
persistency, anonymity, and audibility. Blockchain applications cover a wide range of topics, including cryptocurrencies, financial services, risk management, IoT, and public and social services \cite{zheng2018blockchain}.
The main components of blockchain are discussed in this section, and how each of these components contributes to the technology is detailed.
\subsection{Distributed Ledger}
\begin{figure}[h]
\centering
  \includegraphics[scale = 0.47]{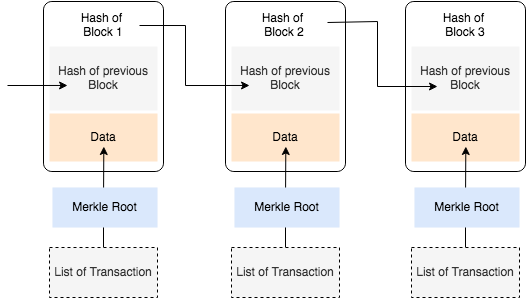}
  \caption{Blockchain Structure.}
  \label{blockchainstructure}
\end{figure}

Blockchain is the concept of a distributed ledger maintained by a peer-to-peer network in a trustless environment. As shown in \figurename{\ref{blockchainstructure}}, Its data structure consists of bundled data chunks called blocks; Blocks are recorded in the blockchain with exact ordering. Briefly, a block contains a set of transactions (exchange and transfer of information); a reference to the preceding block that identifies the block's place in the blockchain; an authenticated data structure (Merkle tree) to ensure block integrity.
\subsection{Peer to Peer Network}
Blockchain uses peer-to-peer networking to achieve the property of decentralized networks with no breaking points. There are two types of nodes in the blockchain network: 1) light node and 2) full node. Full nodes have the public ledger and use blockchain network services to validate new transactions and publish a new block. Light nodes communicate with the blockchain while relying on full nodes to provide them with the necessary information, and as they do not store a copy of the chain, they only query the current status for which block is last and broadcast transactions for processing.Depending on the blockchain's applications, blockchain is divided into public, private, and consortium. A public blockchain is an open, unlicensed structure where anyone can connect to the network and participate in the consensus protocol without having permission or trust from other network nodes and having a copy of the blockchain. Private blockchain has a license-based structure and contains rules. These rules specify who has the right to enter the blockchain network and play a role in its expansion. Finally, a consortium blockchain is a type of blockchain that combines public and private blockchain components.
\subsection{Consensus algorithm}
A blockchain system is, fundamentally, a distributed system that relies on a consensus algorithm. A consensus algorithm is the core component that performs tasks such as deciding whether a transaction has the authority to validate and store on a distributed ledger, selecting nodes to manage the ledger, and ensuring network information integration. Different consensus algorithms have different advantages and limitations. Table \ref{table1} gives a comparison between different consensus algorithms \cite{puthal2018proof,zheng2017overview,bach2018comparative}. 
\begin{table}[tp]
\caption{\label{tab:table-name}A comparison of different consensus algorithms of blockchain.}\label{table1}
\small
\centering
\begin{tabular}{ |p{3.8cm}|p{1.6cm}|p{1.6cm}|p{1.8cm}|  }
 \hline
 \multicolumn{4}{|c|}{Consensus Algorithm} \\
 \hline\hline
 Feature&PoA&PoS&PoW\\
 \hline
 Blockchain type& private&Public and private&Public and private\\
  \hline
 Node identity management&   open  & open   &permissioned\\
 \hline
 Energy consumption&   low  & high   &high\\
  \hline
 Calculation requirement&   low  & high   &high\\
  \hline
 Throughput    &high & low&  low\\
  \hline
 Delay&   low  & high   &high\\
 \hline
\end{tabular}
\end{table}
\subsubsection{PoW}
The proof of work (PoW) consensus algorithm is the most widely deployed consensus algorithm in existing blockchains. PoW is used for establishing a consensus in the public blockchain. PoW is based on a challenge-response method, where the nodes (miners) need to compete by solving certain computational and complex cryptographic problems to add a block onto the chain; This mathematical problem is not easily solved and requires a lot of electricity, cost, effort, and processing power to solve it. For this reason, the probability of finding a solution corresponds with the processing power which is energy intensive. When a miner finds the answer, it announces the solution to the whole network and receives its reward, which is some digital currency, according to the network protocol \cite{gervais2016security}. Bitcoin blockchain employs PoW as the consensus mechanism \cite{nakamoto2019bitcoin}.
\subsubsection{PoS}
Proof of Stake (PoS) has been developed to overcome the constraints (cost and inefficient resource usage) of the PoW algorithm. The core idea of PoS evolves around the concept that the nodes who want to participate in the block creation process must first prove that they own a specific number of coins called a stake. Furthermore, a certain amount of their coins must be locked into an escrow account. The stake serves as an assurance that they will follow the protocol's guidelines. As the stakeholder escrows its stake, it implicitly becomes a member of an exclusive group. Only members of this select group are permitted to participate in the creation of a block. In a PoW, the more processing power you have, the better your chances of creating a new block, but in a PoS algorithm, the more stake you have, the better your chances of creating a new block \cite{saleh2021blockchain}.
\subsubsection{PoA}
In 2017, Ethereum co-founder and former CTO Gavin Wood proposed the Proof of Authority (PoA) as a replacement for the PoW. PoA is a Byzantine Fault Tolerant (BFT) consensus algorithm designed for private blockchains with a limited number of authenticated members (validators). Validators are a group of trusted entities who act as moderators of the system. The validators are responsible for collecting transactions, creating and adding the blocks into the chain. The PoA consensus algorithm is divided into rounds, with each round allowing a validator to submit a block \cite{de2018pbft}.In its corresponding round, a validator proposed a block. The other validators verify the proposed block and add the block to their local copy of the blockchain if it is genuine.\\
\subsection{Smart Contracts }
A smart contract is a self-executing contract consisting of lines of code that establish a set of rules for an application on the blockchain. A smart contract keeps track of the state of an application, and each valid transaction updates the smart contract's state on the blockchain that corresponds to it. Because a smart contract is put on a blockchain, it ensures that the contract's proceedings are transparent without the possibility of changing its rules and any ambiguity in its provisions. Smart contracts scripts allow us to redefine how radically interactions between transacting parties on an IoT network can be set up and automated without the need for intermediaries and third parties \cite{aggarwal2021blockchain}.
\section{Related Work}
In this section, firstly, there exist instances of blockchain's application to IoT, then we focus specifically on the articles related to using blockchain to decentralizing IoT, then we review literature in blockchain PoA consensus algorithm in IoT. 
\subsection{Blockchain in IoT applications} 
Many IoT applications now adopt blockchain for various purposes such as secure data storage, device authentication, access management, privacy enhancement, third party deletion, secure communication between devices \cite{tariq2019security}, and digital payment \cite{lao2020survey}. One application of blockchain in IoT is \cite{dwivedi2019optimized}, where a novel of modified blockchain models suitable for healthcare things have been proposed. It uses blockchain to secure the management and analysis of healthcare big data. Likewise, in \cite{lee2017blockchain}, a blockchain-based system has been proposed to manage the firmware update of IoT devices. Authors in \cite{samuel2019blockchain} exploit blockchain to store access control data to support users' security and privacy in smart grids. The authors of \cite{xu2019blockchain} propose a blockchain-based fair nonrepudiation service provisioning scheme for industrial IoT (IIoT) scenarios in which the blockchain is used as a service publisher and an evidence recorder to eliminate requirements of trusted third parties or unacceptable overheads. \cite{rehman2019cloud} proposes a distributed cloud architecture based on the blockchain technique, which provides low-cost, secure, and on-demand access to the most competitive computing infrastructures in the IoT network and secure provisioning and data sharing systems. The work in \cite{dorri2017blockchain} proposes a framework that relies on hierarchical structure and distributes trust to maintain blockchain security and privacy while making it more suitable for IoT's specific requirements.
\subsection{Blockchain in decentralizing IoT} 
The works mentioned earlier use blockchain to either execute smart contracts or perform application-specific tasks but not decentralize IoT architecture.
Hybrid-IoT, the platform designed for decentralizing IoT architecture in \cite{sagirlar2018hybrid}, exploits both PoW blockchains and BFT consensus algorithm; also, in this article, they define a set of integration metrics and sweet spot guidelines designing decentralized IoT architecture using blockchain.
In \cite{singh2019managing}, blockchain is used in the end-to-end decentralized infrastructure of smart homes to store information and transactions without the need for intermediaries, guarantee the availability of services. It also uses smart contracts for managing appliances.
Authors in \cite{uddin2020blockchain} propose a Blockchain leveraged decentralized eHealth architecture to ensure reliable, secure, and private communication without the need to trust third parties.
Another work by \cite{al2020blockchain} introduces a decentralized composition system solution for IoT based on blockchain for complex multimedia service delivery to cloud subscribers. To authenticate and offer composite services, the proposed work dynamically builds user-defined services without requiring any intermediary service or network provider entities.
\par The number of researches on the use of blockchain in the IoT has been increasing in recent years, and significant achievements have been made, but there are still issues that remain hidden from researchers, issues such as energy managment, efficient use of resources, .
\subsection{PoA blockchain in IoT} 
 In IoT blockchain applications, we have focused on decentralizing the IoT architecture, which means eliminating dependencies on central management for storage and processing. One of the challenges of decentralizing IoT architecture is the energy management of nodes. The more nodes interact in the network, the more critical it becomes to find a solution to this challenge. Inefficient use of available resources is one of the reasons for energy loss. Given the high volume and speed of data production and processing in the IoT, another challenge is to increase the volume of traffic and information density in the system, which causes delays and reduces system performance. Algorithms designed for inter-node processing must be scalable according to demand and workload and manage the large amount of information generated by IoT devices without slowing down. Not only must they be prepared to scale, but they must be able to reduce their scale as workload decreases to avoid spending extra. System reliability means that system components can be replaced in the event of a breakdown without damaging the original data and applications. The accessibility of services is also among the goals that should be considered.
\par Depending on the range of blockchain applications in the IoT, there are different consensus algorithms to implement. In recent articles, the PoA consensus algorithm, which has deficient computational power, low block generation delay, high throughput, and low power consumption, has been considered as a suitable algorithm for the IoT system \cite{puthal2018proof}. Privacy and access management can also be easily implemented because the algorithm runs on private blockchains. The work by \cite{singh2019managing} implement and evaluate a blockchain-based approach using PoA as the consensus mechanism for managing appliances in smart homes. In another work, \cite{lone2020reputation}, a reputation-driven dynamic access control framework for IoT based on PoA Blockchain is provided. Authors in \cite{hakiri2020blockchain} introduced the design of a novel Blockchain-based IoT network architecture to secure IoT networks with a PoA consensus algorithm to detect and report suspected IoT nodes and mitigate malicious traffic. The authors in \cite{qazi2020proof} are proposing PoA to implement blockchain for health care data interconnectivity, interoperability, and data sharing arises. \cite{javed2020blockchain} a blockchain-based secure data sharing mechanism is proposed for VehicularNetworks, and a PoA consensus mechanism is used to validate the transactions. The work in \cite{al2021automated} offers IoT-based blockchain architecture, including five layers: Things, gateway, Fog, Cloud, and application for healthcare to enhance data sharing in a decentralized manner; it also employs PoA due to it is most lightweight and suitable for IoT application due to energy-saving and its election process.

\par Proposed frameworks for using blockchain in the IoT In previous research, the PoA consensus algorithm is presented as a turned base selection (TBS) to select a validator, which does not support resource efficiency. In this method, a validator is selected in each round to confirm transactions and add a block to the blockchain. If the node fails and is not available at the specified time, the specified service cannot be guaranteed. Furthermore, regardless of the capacity of that node, in terms of productivity, tasks are assigned to that node, which makes the pressure on a node more when the network load is high, so that node might not be the best choice to perform tasks and needs more processing, which will increase energy consumption. It also improves response time, resulting in network congestion and latency. This article discusses methods in using blockchain to decentralize the IoT architecture that none of the previous techniques has considered. We will introduce these methods in detail in the following.
\section{System Architecture }
\begin{figure}[h]
\centering
  \includegraphics[scale = 0.2]{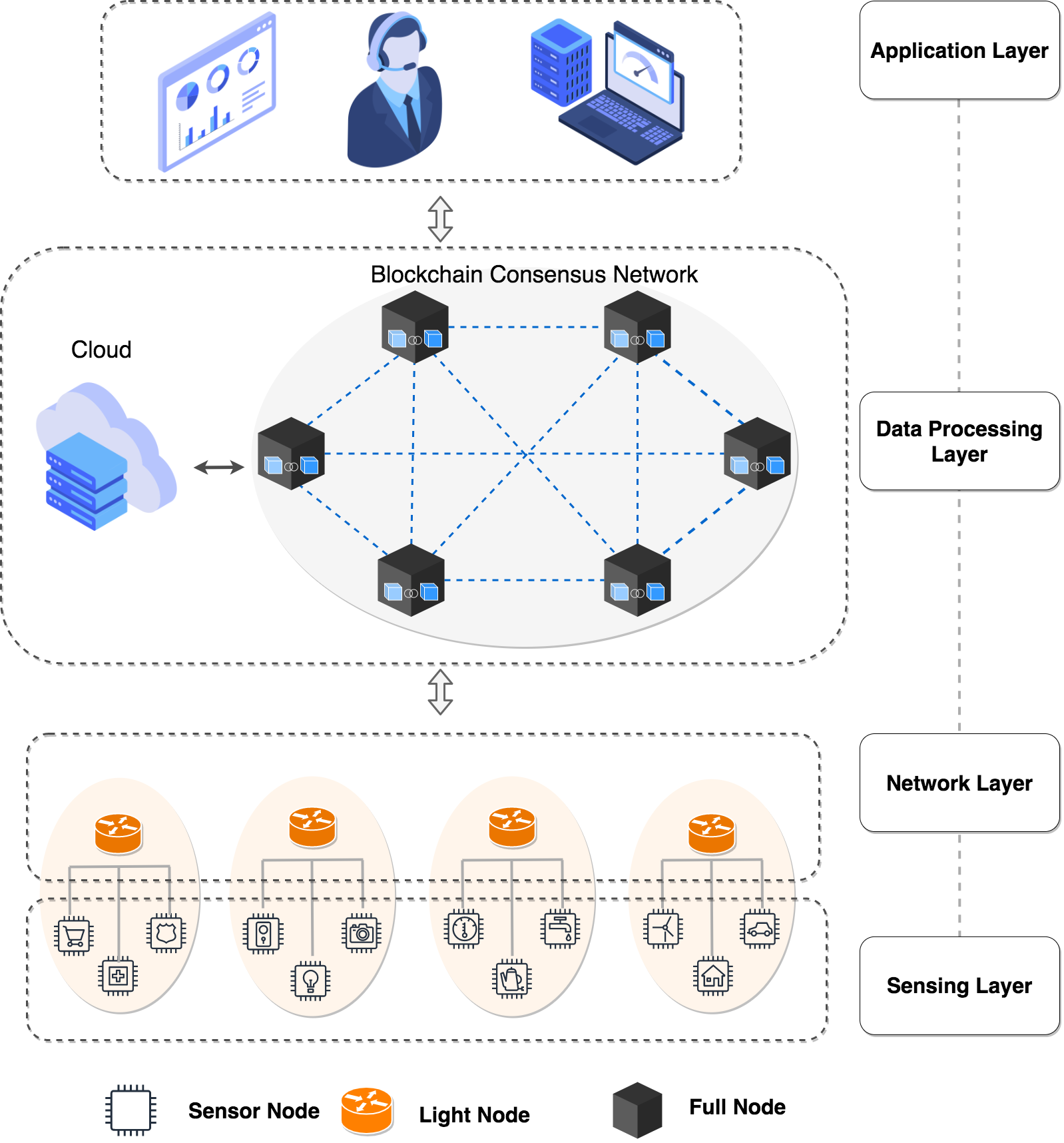}
  \caption{Proposed system Architecture.}
  \label{systemarchitecture}
\end{figure}

An overview of the proposed framework at four-layer IoT architecture is shown in \figurename{\ref{systemarchitecture}}. In IoT blockchain applications, we have concentrated on decentralizing the IoT architecture, which implies reducing dependency on central administration for storage and processing. Advantages include reduced data transmitted to the cloud for analysis and processing, increased accessibility owing to increased system stability, reduced latency, real-time support, increased scalability and flexibility, and privacy and information protection in the event of a security breach.
\par To clarify this issue, we present a blockchain-based smart home scenario. The smart home consists of appliances and devices equipped with computing and communication technologies and communicates with each other to meet the needs of residents. Blockchain can store important transactions between devices and ensure the accuracy, availability, and security of data. Using smart contracts makes it possible to provide a situation where there is no need for a central controller to make decisions. Each device can make the necessary decisions according to the instructions defined in its smart contracts. Each device in the smart home is governed by a smart contract that defines the data, status, and communication instructions of this device with other devices.
\par
\begin{figure}[t]
\centering
  \includegraphics[scale = 0.32]{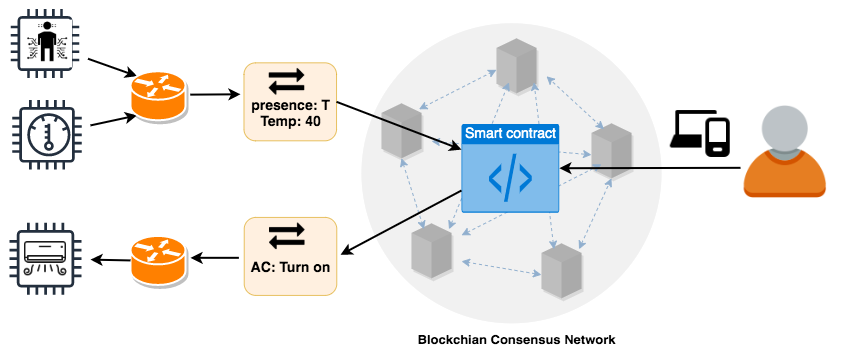}
  \caption{Smart Home Scenario.}
  \label{smarthome}
\end{figure}

 For example, as indicated in \figurename{\ref{smarthome}}, a temperature sensor, the temperature it senses from the environment of a room, is 30 degrees Celsius. On the other hand, a motion sensor detects the presence of a person in the room. A smart contract is formed on the blockchain that controls the air conditioner of that room in various conditions, and instructions are written on it. This smart contract defines the permission to access the data and the status of the temperature sensor and motion sensor, and by receiving the obtained results, it can take appropriate action. For example, if the temperature exceeds the intended threshold, the smart contract updates its status and commands the smart air conditioning system to be turned on. Residents and landlords can interact with this smart contract via a mobile or web-based app by changing the temperature threshold level. These actions are performed as a transaction in the network and are stored in the blockchain.
But in a centralized architecture, these communications take place through a centralized cloud institution.
\par There are four primary components to this framework.
\subsection{IoT devices}
An IoT ecosystem consists of smart devices and sensors that use embedded systems, such as processors, sensors, and communication hardware, to collect, send and act on data they acquire from their environments. IoT sensors share data they collect by connecting to an IoT gateway which can then be transferred to processing units to be analyzed or process. IoT devices have a wide range of capabilities, and their functions should reflect their capacity. As a result, in our proposed blockchain framework, three types of devices are defined.
 \par {\bf Sensor nodes:}
IoT sensor network nodes have insignificant capabilities and can only act as sensors to generate data whose raw data is not stored in the blockchain if it is not critical. The power source of these nodes is generally a limited battery.
\par {\bf  Light nodes:}
IoT devices with limited computing power play the role of light nodes in the blockchain, like Node MCU. These nodes are responsible for simple tasks such as combining data and sending transactions to the consensus network. These nodes do not have the full version of the blockchain and can only store blocks head to confirm the blockchain. These nodes also receive and process sensor node data because they have a higher power source. They filter the raw data generated by the sensors and devices and extract valuable data from it.
\par {\bf Full nodes:}
These nodes are IoT devices that have high computing power and capacity for complex operations, like Raspberry pi3. Their main tasks include maintaining consensus among blockchain nodes, enforcing smart contracts, and approving transactions. They also store a complete copy or part of the blockchain general ledger. The manager stores all or part of the general ledger according to the nodes' energy considerations and storage space. These nodes are also called archive nodes. Also, when a network needs to make a decision, the full nodes vote on the proposals, and if more than 51\% agree, the proposal is accepted and implemented.
\subsection{Blockchain consensus Network}
This network includes blockchain validators (full nodes) that monitor transactions sent by the IoT wireless sensor network. They also execute smart contracts and maintain a distributed general ledger that includes transactions.
\subsection{Cloud}
The cloud is intended as an intellectual position to access powerful computing resources for heavy processing and gain knowledge from the information obtained.
\subsection{Application}
This part connects the system to the user and includes software through which the user can monitor and control it.
\par
\subsection{Proposed methods}
In the following, we discuss the innovations and methods proposed to improve and control the consensus network to increase productivity and efficiency.
\subsubsection{Clustering}
Some characteristics of wireless sensor networks, such as limited power supply, low bandwidth, and inadequate memory capacities, make the network vulnerable. These networks' limited battery life has made energy consumption in wireless networks one of the main challenges. One way to reduce energy consumption and increase scalability in sensors is clustering. In this approach, nodes are organized into a cluster so that high-energy nodes are used in the light role to process and send data. Clusters in wireless sensor networks each contain a main node called a header and several sub-nodes as members. The choice of the header should be such that it distributes energy consumption over the entire network. Cluster size control based on the sensor nodes' CPU power, the correct timing can significantly reduce power consumption and improve service quality. Sensor nodes send their data alternately to their header (light node). Head nodes collect and process data and send a transaction over the consensus network if a transaction needs to be created. The full nodes receive these transactions and check their validity. Once the transaction is approved, permission is granted to execute the transaction or register it in the blockchain.Algorithm complexity, message overhead, and transmission delay are some technical challenges of clustering, but clustering helps collaborate and prolongs the overall network lifetime \cite{cui2019optimal}.
\subsubsection{Virtualization}
\begin{figure}[h]
\centering
  \includegraphics[scale = 0.9]{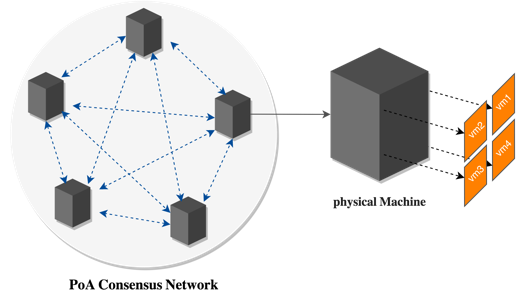}
  \caption{Virtualization in consensus network.}
  \label{bs}
\end{figure}

Virtualization applies better resource allocation, which reduces costs and energy consumption, leading to a greener structure. Also, scalability will increase with the ability to share services. Better and easier management of services and increased security are other advantages. Because of the benefits mentioned, we have also used virtualization in our proposed blockchain-based IoT architecture. Virtual machines are separate and isolated software on which the blockchain application is located. As shown in \figurename{\ref{bs}}, virtual machines run on the physical machine of the blockchain consensus network validator. These virtual machines are responsible for verifying transactions and executing smart contracts. Virtualization has also been proposed as Ethereum Virtual Machine (EVM). The cost of the implementation is the main challenge of virtualization. Still, virtualization, if utilized efficiently with proper planning we can extract the maximum output of this helpful tool in blockchain consensus network.
\subsubsection{Weight Based Selection (WBS)}
Tasks are assigned to full nodes physical machines, and each physical machine is subdivided into several virtual machines that can accomplish the tasks simultaneously. As mentioned in the previous works, Turned Based Selection (TBS) has been used in the PoA consensus network, and the challenge of inefficient use of available resources, including reasons for energy loss, latency, scalability, and reliability. Due to the dynamic requests, such as the variability of transaction arrival time and the number of requests in the wireless sensor network, it is proper to use solutions to distribute tasks between validators. Our goal is to provide a modified version of the PoA consensus algorithm for the blockchain-based IoT network to maximize network potential and assign tasks based on service requirements. We have proposed a method of weight Based Selection (WBS). To make the best use of consensus network nodes, we select them based on the weight of attractiveness. This criterion is to select the most appropriate validator. We use \eqref{attra} to calculate attractiveness.
\begin{equation}
Attractiveness_{PM} =IBscore_{PM}*LoadFraction_{VM,PM}
\label{attra}
\end{equation}
In this regard,$IBscore_{PM}$is the processor efficiency of each physical machine validation and is obtained from \eqref{ibscore}.
\begin{equation}
IBscore_{PM}(u,T_{upper})=  e^{(u-T_{upper})}
\label{ibscore}
\end{equation}
$u$ is the CPU and RAM efficiency, and $T_{upper}$ is the physical validation machine's productivity threshold. If $u$ is less than $T_{upper}$ it will be negative, and the $IBscore_{PM}$ value will decrease. The lower the $IBscore_{PM}$ value, the less obligation there is on the physical machine.\\
Also, the $LoadFraction_{VM,PM}$ value is a fraction of the physical machine's available resources, which is allocated to the virtual machine and is obtained from \eqref{loadfraction}.
\begin{equation}
LoadFraction_{VM,PM} = \frac{S_{VM}}{S_{PM}}
\label{loadfraction}
\end{equation}
In \eqref{loadfraction}, the numerator, ${S_{VM}}$, shows the processor assigned to the virtual machine, and the denominator, ${S_{PM}}$, shows the CPU amount remaining in the physical machine (If there is no CPU remaining in the physical machine, the system should skip this physical machine). The lower the LoadFraction of a physical machine, the more available the physical machine is and the less likely it is to become overloaded. 
\par Therefore, it can be concluded that the validator is appropriate if the multiplication of $IBscore_{PM}$ and $LoadFraction_{VM,PM}$ , which is attractiveness weight, is less. Accordingly, in the next step, the validator's virtual machines are sorted according to the weight, and based on this, the virtual machines with the minimum weight select to confirm the transactions and add a new block.
\par Our method overhead is a computational overhead that includes WBS calculations to give weight to the validators. For this purpose, we first check the TBS method by a controller (it could be a smart contract on a blockchain) before performing WBS calculations. If we do not have enough processing power in the TBS validator at the scheduled time to perform the tasks, we start WBS calculations. The generated overhead has the significant benefits of reducing response time, increasing scalability, and reducing energy consumption.
\par The WBS algorithm in the proposed framework is as follows.
\begin{algorithm}[t]
\scriptsize
\SetAlgoLined
initialization: transactions are sent over the WSN-IOT\;
\textbf{Step1}, Set Validator PM (Physical machine) and related VM (Virtual    machine) to Task with TBS algorithm (Round Robin);\\

  \eIf{VM from PM can handle the task}{
   the operation is done;
   }{ \textbf{Step2}, WBS calculation will be processed;\\
   \For{each active Validator}{
    calculate$LoadFraction_{VM,PM} = \frac{S_{VM}}{S_{PM}}$\;
    calculate$IBscore_{PM}(u,T_{upper})=  e^{(u-T_{upper})}$\;
    calculate$Attractiveness_{PM} =IBscore_{PM}*LoadFraction_{VM,PM}$\;
    }
Sorting results from WBS, Computing and select PM with related 
VM for handling the task;
  }
 \caption{procedure of selecting validator.}
\end{algorithm}
\section{Simulation}
\begin{figure}[h]
\centering
  \includegraphics[scale = 0.38]{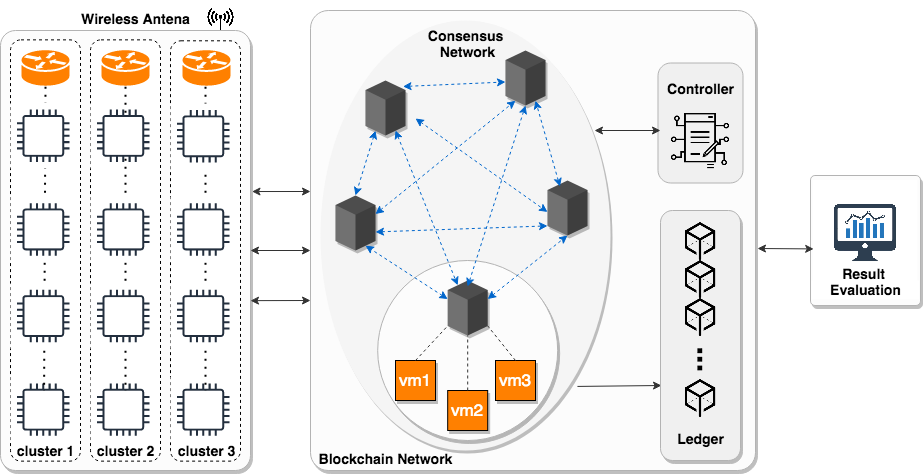}
  \caption{The scheme of simulation's sections.}
  \label{simulation}
\end{figure}

To evaluate the proposed method, we simulated it using the NS2, a widely used IoT simulator. The proposed scheme for simulation and evaluation of the proposed framework is shown in \figurename{\ref{simulation}}. This plan's implementation includes parts of the wireless sensor network, blockchain network, and output processing to obtain evaluation criteria.
\par We used the MannaSim framework in NS2 to implement the IoT wireless sensor network (WSN) and clustering. This is a modular framework for simulating the WSN for design, development, and analysis in various applications \cite{pereira2015mannasim}. Two types of nodes, common node and cluster head, whose classes are defined in MannaSim, are considered in this network.
\par We have used the LEACH algorithm for clustering \cite{xiangning2007improvement}, which is suitable for a WSN to reduce energy consumption and increase the network's lifetime; it has been implemented in MannaSim. This algorithm is one of the most basic and popular hierarchical clustering algorithms and has two establishment and stability phases. This algorithm's establishment phase is that the nodes are identified first, and then the nodes join the nearest cluster head, and clusters are formed. Following that, the stability phase is implemented. All sensor nodes collect the sensed information from the environment and then send it to their head. This reduces network traffic and energy consumption.

\par The blockchain network consists of physical validation machines that include several virtual machines. Their specifications are listed in table \ref{configurations}. To implement virtual machines, we used the Green-Cloud tool in NS2 \cite{kliazovich2012greencloud}. This tool models their physical machines, virtual machines, communication links, and power consumption and can be used to develop solutions for resource allocation, task scheduling, and optimization of communication protocols.

\begin{table}[!htb]
\centering
\caption{physical machine and virtual machine configurations.}
\label{configurations}
\scalebox{0.8}{
 \begin{tabular}{|c|c|c|} 
 \hline
 Type	&CPU	&MEM (GB)\\ [0.5ex] 
 \hline\hline
 Virtual machine&	1 (1 cores x 1 units)&	1.7\\ 
  \hline
 Physical machine&	4 (4 cores x 1 units)&	8\\
 \hline
\end{tabular}}
\end{table}
To implement the turn based selection (TBS) algorithm for assigning tasks, we used the Round Rubin (RR) algorithm in the Green-cloud simulator. This algorithm works by placing the virtual machines in a queue and assigning it to the first virtual machine in the queue each time a request is entered, then moving that virtual machine to the queue's end. We explained the proposed weight-based selection (WBS) algorithm in detail in the previous chapter; It was implemented in C ++. The blockchain public ledger was also implemented in C ++ and used in NS2. Each block contains an index, timestamp, Merkel (a list of transactions), hashes, and hashes of the previous block. With each new block developed by the validator, a new version of the blockchain is sent to all nodes. Table \ref{parameters} shows the simulation parameters.
\begin{table}[!hbt]
\centering
\caption{physical machine and virtual machine configurations.}
\label{parameters}
\scalebox{0.8}{
 \begin{tabular}{|c|c|} 
 \hline
 Parameters	&Value\\
 \hline\hline
 Mac channel&802.11 ieee\\ 
  \hline
 Traffic type&CBR\\
 \hline
 Routing protocol&AODV\\
 \hline
 Transfer protocol&TCP\\
 \hline
 Clustering protocol&LEACH\\
 \hline
 Node position&Random way point\\
 \hline
 Queue type&Drop tail\\
 \hline
  Queue length&50\\
  \hline
  Antenna type&OmniAntenna\\
 \hline
 Initial energy common node&3J\\
 \hline
 Initial energy cluster head&5J\\
 \hline
 Initial energy validator&10J\\
 \hline
 Area&100 m\\
 \hline
\end{tabular}}
\end{table}

\section{Performance Evaluation}
To provide a comprehensive evaluation, We have considered the number of validators and IoT devices as variables. Number of nodes is suggested in the article \cite{sagirlar2018hybrid}, in proportion to 7\% of the validators ratio to all devices. Also, IoT devices have different tasks, and various transaction submission periods are defined in them. Therefore, to evaluate our proposed WBS method's performance toward the TBS algorithm, we subjected it to different dissemination intervals of transactions from IoT sensors.
We have used the following metrics for evaluation:
\begin{itemize}
 \item Response time: Refers to the processing time of a transaction, from the time it is created to save in the blockchain.
 \item Throughput: This metric refers to the size of the packet transmitted per second.
 \item Energy consumption: Refers to the energy consumption of the validators for processing WSN IoT transactions and developing a new block.
\end{itemize}
\par
The discussion on the evaluation is as follows:
\subsection{Response time}
\begin{figure}[t]
\centering
  \includegraphics[scale = 0.38]{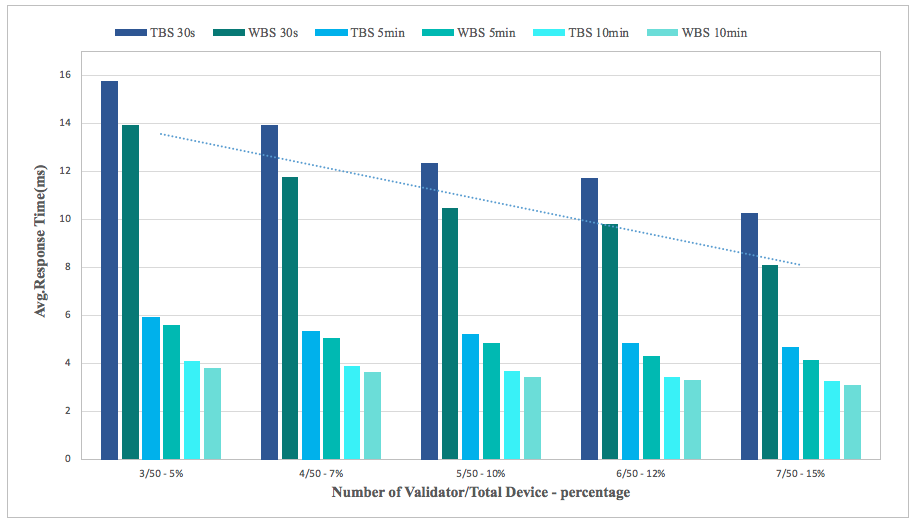}
  \caption{Comparison of WBS and TBS in terms of response time on the different number of validators with different transaction submission periods.}
  \label{res}
\end{figure}

Because the volume and speed of data generated in the IoT are very high, it is desirable to improve network performance by reducing latency and traffic nodes and increasing scalability. Delays in sending and receiving information between nodes can cause significant problems to latency-sensitive applications.
To this end, using the maximum capacity of the network and reducing delays in processing will reduce unwanted traffic load, and is desirable.\\
In one scenario, we estimated the proposed method's response time in transaction dissemination intervals of 30 seconds, 5 minutes, and 10 minutes. We have repeated this evaluation in different consensus network sizes. The results in \figurename{\ref{res}} show that, compared to TBS, our proposed model WBS has improved response time by increasing requests (30-second transaction dissemination interval) and providing 15\% less response time.
\par
\begin{figure}[h!]
\centering
  \includegraphics[scale = 0.41]{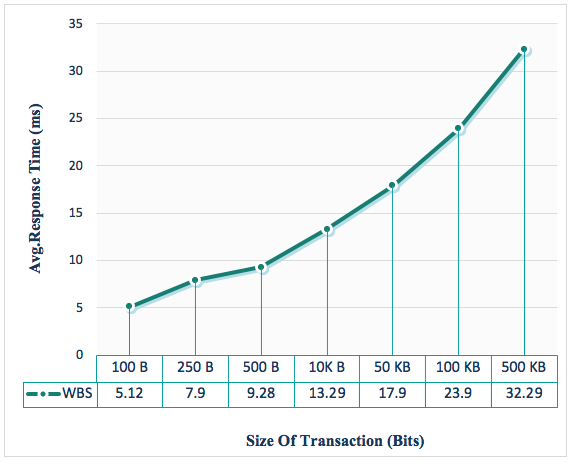}
  \caption{Impact of transaction size on response time.}
  \label{sizeres}
\end{figure}

 As mentioned earlier, different IoT transactions are defined for various applications. To obtain the desired point of our proposed method, we have measured the average response time of transactions with varying sizes of transactions. The results in \figurename{\ref{sizeres}} show that the smaller the transaction size, the shorter the response time. This is in line with our goal of optimizing IoT in blockchain, which also increases scalability. For this purpose, small-sized transactions are proposed in the design of the proposed framework.

\subsection{Energy Consumption}
\begin{figure}[h!]
\centering
  \includegraphics[scale = 0.39]{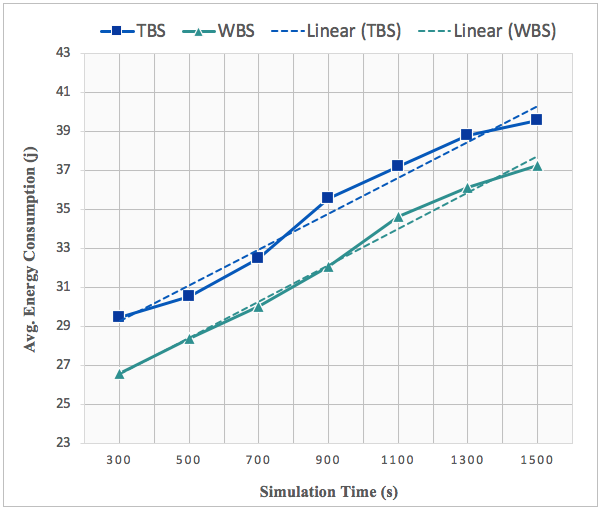}
  \caption{Comparison of WBS and TBS from the point of view of energy consumption in the blockchain consensus network.}
  \label{energy1}
\end{figure}

One of the challenges of IoT decentralization is energy management in nodes. The more nodes interact in the network, the more critical it becomes to find a solution to this challenge. Network nodes may have limited energy storage, so they cannot be assigned large processing volumes. Inefficient use of available resources is one reason for energy loss, so resource management methods with energy efficiency approaches are needed to schedule and allocate resources based on service needs.\\ As energy consumption in the IoT increases, support for green computing has increased. By proposing the WBS method, we have tried to make better use of resources and, as a result, reduce energy consumption and increase network life. \figurename{\ref{energy1}} shows the consensus network energy consumption for WBS and TBS in the scenario with 50 sensor nodes, 5 light nodes (head cluster), and 4 full nodes. The WBS method has reduced energy consumption by 8\% compared to TBS.
\par
\begin{figure}[h!]
\centering
  \includegraphics[scale = 0.38 ]{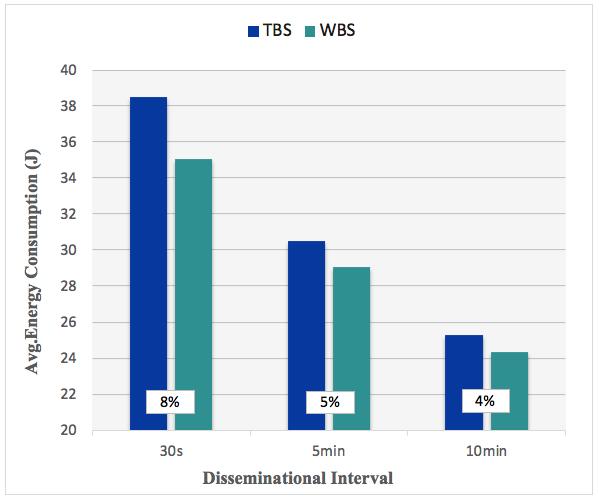}
  \caption{ Comparison of average WBS and TBS energy consumption in different transaction Dissemination Intervals.}
  \label{energyinterval}
\end{figure}

 We also evaluated the amount of energy consumption in different transaction dissemination periods, 30 seconds, 5 minutes, and 10 minutes. The results in \figurename{\ref{energyinterval}} show that the performance of our proposed method, WBS, has better results than TBS in fewer transaction dissemination intervals which involving more requests.

\subsection{Throughput}
\begin{figure}[h!]
\centering
  \includegraphics[scale = 0.39]{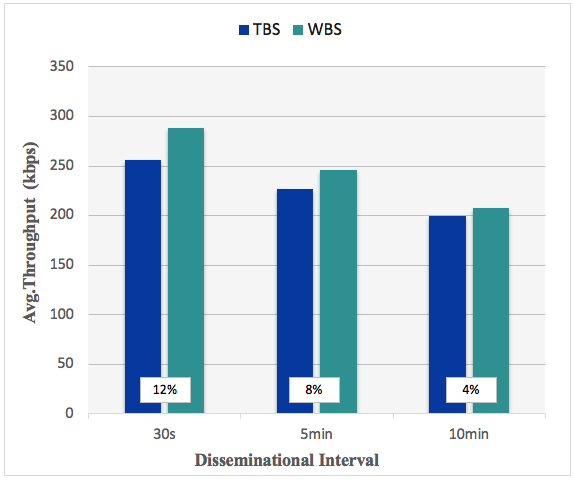}
  \caption{Comparison of WBS and TBS in terms of throughput on different transaction dissemination intervals.}
  \label{throughput}
\end{figure}

High data traffic reduces the operational capacity and efficiency of the network. Furthermore, the high rate of data transmission from device to device on the IoT network can prevent network traffic. For this reason, we have considered this criterion to evaluate our proposed method. \figurename{\ref{throughput}} shows that the throughput of the WBS network in the scenario with 50 sensor nodes, 5 light nodes, and 4 full nodes has increased by 12\% compared to TBS under different IoT wireless sensor network requests. As discussed earlier, we have made progress in throughput due to validator choice with higher processing power and more available resources.
\section{Conclusion}
In this article, we provide solutions to optimize the use of blockchain in the decentralized IoT architecture. This dissertation's primary focus has been to reduce energy consumption, increase scalability, reduce response time, and increase system reliability. To achieve these goals, we used methods such as clustering and virtualization in the framework. We also introduced the Weight Based  Selection (WBS) method in the PoA consensus algorithm to increase system efficiency. By simulation, we compared the performance of our proposed method WBS with the TBS method. The results show that we have improved in reducing energy consumption and response time and increasing throughput. Our method overhead is a computational overhead that includes WBS calculations to give weight to the validators. The overhead generated is worth the significant benefits mentioned.
\par
It is suggested that in future research, to increase the system's productivity, migration methods will be used. In this method, if there is additional overhead in the virtual machines, the tasks will be transferred to other virtual machines to avoid wasting time. Migration is possible both internally and externally.
\bibliography{elsarticle-template}

\end{document}